\documentclass{pasj00}

\begin{document}
\SetRunningHead{Matsunaga et al.}{SiO Maser Sources toward Globular Clusters}
\Received{2004/12/06}
\Accepted{2004/12/22}

\title{SiO Maser Sources toward Globular  Clusters}

\author{Noriyuki Matsunaga,$^1$ Shuji Deguchi,$^2$
Yoshifusa Ita,$^{1,3}$ Toshihiko Tanab\'{e},$^{1}$}
\and
\author{Yoshikazu Nakada$^{1,4}$}
\affil{$^{1}$Institute of Astronomy, School of Science, The University of Tokyo, 
 \\ Osawa 2-21-1, Mitaka, Tokyo 181-0015}
\affil{$^{2}$Nobeyama Radio Observatory, National Astronomical Observatory, 
\\ Minamimaki, Minamisaku, Nagano 384-1305}
\affil{$^{3}$Institute of Space and Astronautical Science, Japan Aerospace Exploration Agency, 
\\ Yoshinodai 3-1-1, Sagamihara, Kanagawa 229-8510}
\affil{$^{4}$Kiso Observatory, School of Science, The University of Tokyo, 
\\ Mitake, Kiso, Nagano 397-0101}


\maketitle

\begin{abstract}

We report on the detection of SiO masers in Asymptotic Giant Branch 
variables toward bulge/disk globular clusters.
In five out of six cases,  the radial velocities are compatible
with the optically measured radial velocities of globular clusters
in the assessed uncertainty. Two sources, toward Terzan~5 and Terzan~12,
lie very close to the cluster centers. The objects toward Pal~6 and Terzan~12
have luminosities appropriate to the AGB tip
in globular clusters, while those toward NGC~6171, Pal~10, and Terzan~5 are
brighter than expected. It is suggested that
the latter three may have evolved from
merged binaries,  offering a test for binary-evolution scenarios 
in globular clusters, if the membership is approved.
\end{abstract}

\section{Introduction\label{sec:Intro}}

Globular clusters (GCs, hereafter) are well-studied stellar systems containing
the oldest stars in the Galaxy, and
much of our knowledge of stellar dynamics and stellar evolution rests on them.
With an age of about 10 Gyr, low-mass ($\sim$ 1 $M_\odot$) 
stars are now on the AGB phase in GCs.
While younger, more massive, AGB stars often turn into infrared stars
with thick dust shells due to heavy mass loss, the low-mass stars
found in GCs do not appear to evolve in this manner.
Although several red giants in GCs show
small mid-infrared (MIR) excesses that indicate weak mass loss
(\cite{Origlia-2002}; Ramdani, Jorissen 2001),
they bear little resemblance to the dust-enshrouded stars found
in young Magellanic clusters \citep{Tanabe-1997}.
Maser emission is another characteristic often seen
in such objects (see \cite{Habing-1996}). However,
no maser sources have ever been found in GCs
(Knapp, Kerr 1973; \cite{Bowers-1979}; Cohen, Malkan 1979;
Dickey, Malkan 1980). A possible exception is the OH maser source
V720 Oph in NGC~6171 (Frail, Beasley 1994).
\citet{Feast-1994} suggested two possibilities for this object:
to be a foreground star or to be an evolutionary product of a merged binary.
Whereas previous surveys searched for OH and/or H$_2$O masers,
no SiO maser detection has yet been reported
(for negative results with the VLA, see the Web page
\footnote{(http://www.computing.edu.au/\~{ }bvk/astronomy/HET608/
 project/).}).
In this paper, we report on a result of SiO maser detections in AGB variables,
possibly associated with GCs.

\section{Observations and Results\label{sec:Obs}}

Targets were selected from our program of near infrared (NIR)
observations of more than 140 GCs taken from the known galactic total of 150.
NIR monitoring observations started in 2002 March using an array camera 
(SIRIUS) attached to the 1.4 m IRSF telescope
at the South African Astronomical Observatory
[see \citet{Nagashima-1999} and \citet{Nagayama-2003} for
instrumental details]. During the course of this work,
we found several new red variables with $(J-K)_0 \geq 3.0$ mag.
These stars are redder than
the samples with small infrared excesses, studied by \citet{Origlia-2002}
or \citet{Ramdani-2001}, at least in $(J-K)_0$,
so they are more promising candidates for SiO maser emission.

\begin{figure*}
\begin{center}
 \FigureFile(150mm,200mm){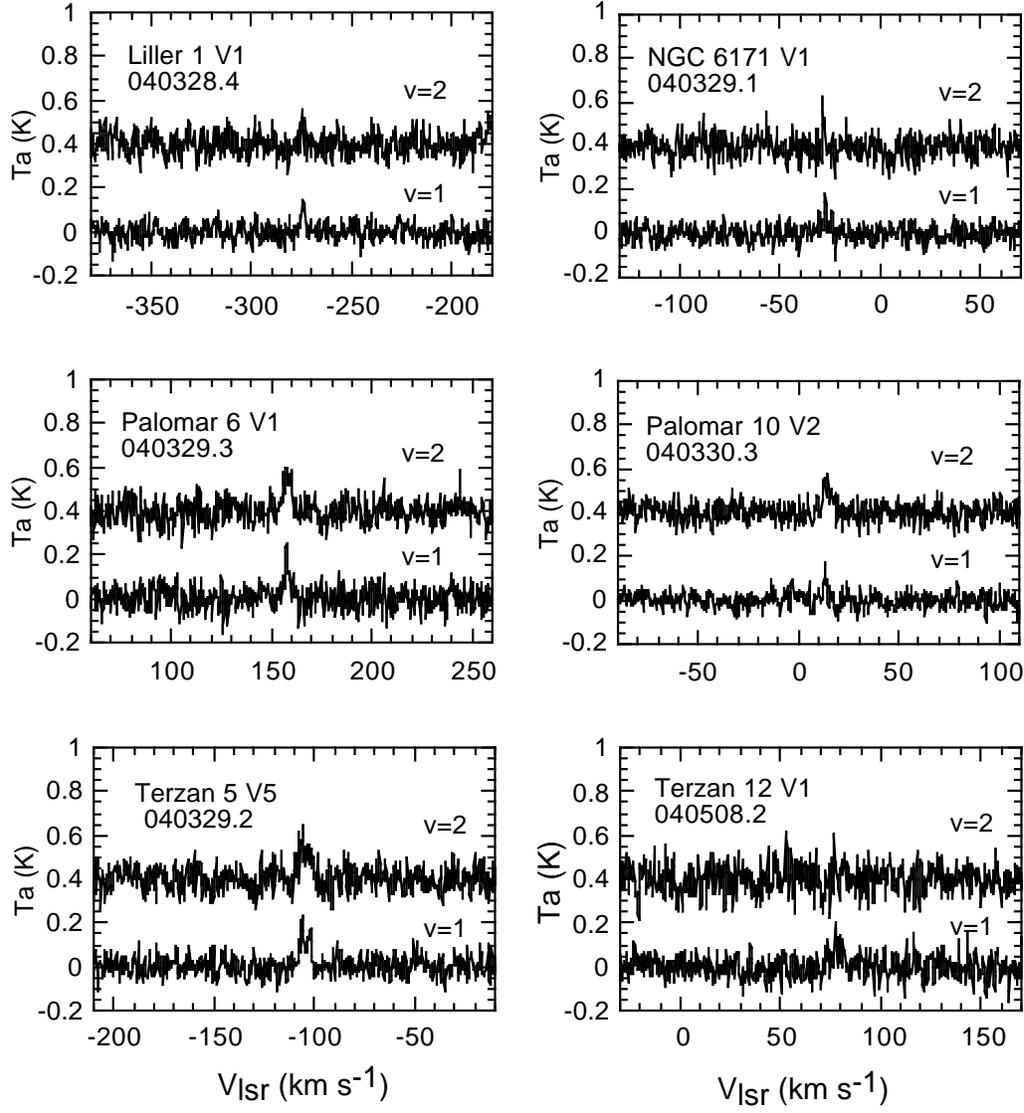}
\end{center}
\caption{Spectra of the SiO $J=1$--$0, v=2$ (top) and $v=1$ line (bottom).
The observation dates are listed in the format YYMMDD.D for all objects.
\label{fig:1}
}
\end{figure*}
For thirty two stars, the SiO maser lines ($J=1$--$0, v=1$ and 2) 
at 43.122 and 42.821 GHz  \citep{Lovas-1992}, respectively,
were searched for with the Nobeyama 45 m
radio telescope during 2004 March 28--30 and on May 8.
The half-power beam width was about 40$^{\prime\prime}$ at 43 GHz. 
We used a cooled SIS receiver ($T_{\rm sys} \sim$ 180 -- 250 K)
and acousto-optical spectrometers with high (40 kHz; AOS-H) and
low (250 kHz; AOS-W) resolutions having 2048 channels each.
The spectrometer arrays
covered  velocity ranges of $\pm 390$ km s$^{-1}$ and  $\pm 800$ km s$^{-1}$
in AOS-H and AOS-W, respectively.
The conversion factor of the antenna temperature to the flux density was
$\sim 2.9$ Jy K$^{-1}$ .

We detected SiO maser emissions from six red variables toward six GCs (among 32 surveyed stars),
mostly in the direction of the galactic bulge.
For the previously known variable stars in NGC~6171 and Pal~10,
the variable IDs (V1 and V2, respectively) were taken from 
the catalog of variable stars in GCs by \citet{Clement-2001}.
NGC~6171 V1 is V720 Oph, mentioned in section \ref{sec:Intro}.
The other four variables were newly found from our NIR observations,
and named according to Clement's numbering system
(namely Liller~1 V1, Pal~6 V1, Terzan~5 V5, and Terzan~12 V1).
Lines of both $v=1$ and $v=2$ were detected, except in the case of Terzan~12 V1,
for which only $v=1$ was seen.
Their spectra from AOS-H are shown in figure \ref{fig:1}.
The detection criteria used by \citet{Izumiura-1999}
were satisfied by these lines, except for marginal detections of the $v=2$ lines
in Liller~1 V1 and NGC~6171 V1. The integration time for each object was about
30 minutes,  similar to those made in the past SiO surveys
(\cite{Izumiura-1999}; Deguchi et al. 2000ab).
The detections were confirmed in AOS-W spectra in all cases.
The observational properties of the detected objects
are summarized in table \ref{tab:1}, which contains the source positions,
angular separations from the cluster center ($r_{\rm SiO}$),  SiO transitions, 
radial velocities ($V_{\rm LSR}$), peak antenna temperatures ($T_{\rm a}$),
integrated intensities,
and rms noise levels for both transitions from the AOS-H spectra.
Table \ref{tab:2} lists the properties of the GCs. 
They were mostly taken from a compilation by \citet{Harris-1996}.
In the table, $l$ and $b$ are the galactic coordinates of the cluster center,
$V_{\rm LSR}^{\rm GC}$ is the cluster velocity with respect to
the Local Standard of Rest (with the escape velocity $v_{\rm h}$
at half radius between parentheses),\footnote{Taken from the table
of escape velocities used for \citet{gne02}, which is available at
http://www-astronomy.mps.ohio-state.edu/$\sim$ognedin/gc/ . }  
$r_{\rm h}$ and $r_{\rm t}$ are the half-mass radius
and the tidal radius of the cluster respectively, and [Fe/H] is the metallicity.
In addition, we give DM (the distance modulus of each clusters) and $E(B-V)$
(the reddening) used to calculate
the absolute bolometric magnitudes in subsection \ref{sec:Mbol},
and $N_{\rm 2M}$ and $P_{\rm contm}^{<r_{\rm SiO}}$,
the number of NIR SiO candidates
within 34$^\prime$ of the cluster and the contamination probability,
which are discussed in subsection \ref{sec:bulge}.

\begin{table*}
  \caption{SiO maser properties.}\label{tab:1}
 \begin{center}
  \begin{tabular}{lccccrccr}
  \hline\hline
Object & RA(J2000) & Dec(J2000) & $r_{\rm SiO}$ & Transition & $V_{\rm LSR}$ & 
Peak $T_{\rm a}$ & Integ. int. &  rms  \\
\hline
  &  h ~ m ~ s & $^\circ ~~' ~~ ''$  & ($'$) & ($J, v$)  & (km~s$^{-1})$ & (K) &
(K~km~s$^{-1}$) & (K) \\
Liller~1 V1 & $17~33~35.2$ & $-33~22~01.8$
 & 2.59 & 1--0, ~1& $-$275.1  & 0.143 & 0.337 & 0.039 \\ 
 &  & & & 1--0, ~2 &  $-$275.1& 0.162 & 0.126 & 0.049   \\ 
NGC~6171 V1 & $16~32~24.6$ & $-13~12~01.3$
 & 8.98 & 1--0, ~1& $-$28.1 & 0.185 & 0.224 & 0.035  \\ 
 &  & & & 1--0, ~2 & $-$28.8& 0.230 & 0.201 & 0.048  \\
Pal~6 V1    & $17~43~49.5$ & $-26~15~27.9$
 & 2.67 & 1--0, ~1& 156.8 & 0.253 & 0.604 & 0.049  \\
 &  & & & 1--0, ~2 & 157.2 & 0.229 & 0.739 & 0.045 \\
Pal~10 V2   & $19~17~51.5$ & $+18~34~12.6$
 & 2.51 & 1--0, ~1& 12.7 & 0.193 & 0.332 & 0.033 \\ 
 &  & & & 1--0, ~2 & 14.7 & 0.173 & 0.653 & 0.041 \\ 
Terzan~5 V5 & $17~48~03.4$ & $-24~46~42.0$
 & 0.34 & 1--0, ~1& $-$106.8 & 0.232 & 0.778 & 0.043  \\ 
 &  & & & 1--0, ~2 & $-$105.9& 0.246 & 0.868 & 0.051  \\ 
Terzan~12 V1& $18~12~14.2$ & $-22~43~58.8$
 & 0.64 & 1--0, ~1& 76.9 & 0.208 & 0.621 & 0.045  \\ 
 & & &  & 1--0, ~2 &  ---     &  ---   &  ---   & 0.065        \\
   \hline
  \end{tabular}
  \end{center}
\end{table*}


\begin{table*}
  \caption{Cluster properties.}\label{tab:2}
  \begin{tabular}{lrrrrrccrrr}
  \hline\hline
Cluster& $l$& $b$& $V_{\rm LSR}^{\rm GC}$~($v_{\rm h}$) & $r_{\rm h}$&
$r_{\rm t}$& DM&  $E(B-V)$ & [Fe/H]
&  $N_{\rm 2M}$($N_{\rm MSX}$) & $P_{\rm contm}^{<r_{\rm SiO}}$ \\
\hline
& ($^\circ$)& ($^\circ$)& (km s$^{-1}$)& ($^\prime$)& ($^\prime$)& (mag)& (mag)& 
&  & \\
Liller~1        & 354.84 & $-$0.16 & 61.0~(19.5)           & 0.45 & 12.57    & 14.81 & 3.06 & 0.22       &  80(187) &   0.207 \\
NGC~6171  &   3.37   & 23.01    & $-$20.4~(17.2)     & 2.70 & 17.44     & 14.02 & 0.33 & $-$1.04 &   1(5)$^\dagger$ &   0.034 \\
Pal~6          &   2.09   &  1.78      & 193.7$^*$(20.9)   & 1.06 &  8.36      & 13.82 & 1.46 & $-$1.09 &  52(96) &   0.148 \\
Pal~10        &  52.44  &  2.72     & $-$13.3$^*$(13.6) & 0.99 &  3.08     & 13.79 & 1.66 & $-$0.10 &    5(19) &   0.014 \\
Terzan~5   &   3.84   &  1.69      & $-$82.4~(32.2)      & 0.83 & 13.27     & 14.97 & 2.15 & 0.00       &  64(70) &   0.003 \\
Terzan~12 &   8.36   & $-$2.10 & 106.3$^*~$(7.8)     & 0.84 &  3.10      & 13.33 & 2.06 & $-$0.50  &   34(64) &   0.006 \\
   \hline
   \multicolumn{11}{l}{$^*$ The value is assessed to be uncertain more than 10 km s$^{-1}$ (see  subsection \ref{sec:Position}).} \\
  \multicolumn{11}{l}{$^\dagger$ Evaluated from IRAS PSC and 2MASS with $K<10$ and $H-K>0.6$ (MSX not available at this $b$).}
  \end{tabular}
\end{table*} 


\section{Discussions\label{sec:Discuss}}

\subsection{Position and Velocity\label{sec:Position}}
All six sources are located well within the projected tidal radii, $r_{\rm t}$
of the GCs. Terzan~5 V5 and Terzan~12 V1 lie within the half-mass radii,
$r_{\rm h}$, i.e., very close to the cluster centers within 1$^\prime$.  

Because the radial velocity of SiO maser emission is
known to coincide with that of the central star within a few km s$^{-1}$ \citep{Jewell-1991},
it can be compared directly with the velocity of the cluster.
Except for Liller~1 V1, the SiO radial velocities fall near to the optically
measured velocities of GCs with velocity separations of 8 -- 37 km s$^{-1}$. 
The velocity dispersion within a GC depends on its mass;
about 10 km s$^{-1}$ for the most massive clusters, and
only a few km s$^{-1}$ for small ones (see table 2 of \cite{Mandushev-1991}).  
\citet{gne02} calculated the escape velocities (as well as the velocity dispersions) at half radii and the cluster centers
of 147 GCs, using isotropic King models with  a constant mass-luminosity
ratio ($M/L_{\rm V}=3$); these escape velocities 
at the half radii are listed between the parentheses in the 4th columns
of table 2. For NGC 6171 V1 and Terzan 5 V5,  the observed velocity differences
of  the SiO masers to the GCs are well within the escape velocities of
the globular clusters,
though they are slightly above the one-dimensional velocity dispersions
given in the Gnedin's table.
 For Pal 6 V1 and  Pal 10 V2, the velocity differences are
a factor of about 1.8   larger than the escape velocity,  and 
for Terzan 12 V1, a factor of 4.2  larger.  
However, note that the accuracy of the radial-velocity measurements is
quite poor  for these three clusters.
They were obtained from the average of the radial velocities of a few giants,
which were selected from a sample of a handful of stars, 
excluding objects with a large velocity separation; 
historic examples in velocity measurements of GCs suggest
a typical uncertainty of about
$\pm$20 km s$^{-1}$ (see table 3 of \cite{coe01}). 
For Pal~6, the velocity was derived  from only two 
(table 3 of \cite{Minniti-1995a}) 
or four  giants (\cite{lee04});  
their values differ by about 20 km s$^{-1}$.
For Pal~10 and  Terzan~12, the radial velocities of only two stars were
used \citep{Cote-1999}.

With this caveat, the velocities of the SiO masers are considered not to be
incompatible with those of the clusters, except in the case of Liller~1 V1,
whose membership is rejected.
Better optical velocity measurements of the surrounding giants
would be necessary to establish the memberships of the SiO masers
from the velocity coincidence with greater certainty.
On the other hand, the radial velocity of NGC~6171 has been studied based on
the spectra of a number of giants (Da Costa, Seitzer 1989;
\cite{Suntzeff-1993}; \cite{Piatek-1994}).
\citet{Harris-1996} gives a velocity of $-20.4$ km s$^{-1}$,
obtained by combining their results. This is in reasonable agreement with
the SiO maser velocity ($-$28.4 km~s$^{-1}$), despite the large angular separation ($\sim 9'$) 
from the cluster center.
There have also been proper-motion studies concerning this cluster,
which support the membership (Frail, Beasley 1994).
Also, for Terzan 5, the radial velocity, which was recently derived
using 4 stars (Origlia, Rich 2004),  coincides with the previous value listed
by \citet{Harris-1996},
as well as with the SiO maser velocity within a  tolerable range.
Thus,  NGC~6171 V1 and Terzan 5 V5 
cannot be rejected as being  members of the clusters from
a kinematic point of view.

\subsection{Concentration of Detections to the Galactic Bulge \label{sec:bulge}}
Among the six GCs, five are located in the galactic bulge, the exception
being Pal~10, which is in the galactic disk.
Because of the high density of AGB stars in these areas,
possible contamination by the bulge AGB stars with  SiO masers
in the line of sight must be examined.
The surface number density of SiO maser sources is estimated to be
approximately 10--20 sources per square degree in the galactic bulge;
this number was obtained by averaging the numbers found in the inner bulge
in past surveys covering an area of $25^\circ\times6^\circ$
(Deguchi et al. 2000ab),
corrected (with multiplying by a factor of ten) for limited color ranges
[$0<\log(F_{25}/F_{12})<0.1$] and flux density limits ($F_{12}>2$ Jy) 
of the above-noted observations.
This gives $0.16$ at most for the number of spurious
SiO sources falling within 3$^\prime$ from the center of a cluster in the bulge.
Furthermore, a  probability of contamination by the disk/bulge objects
was evaluated for individual  directions of GCs by counting MSX sources 
with 12 $\mu$m flux density above 1 Jy having a NIR counterpart falling within 5$''$ with $K<10$ and $H-K>0.8$
 in the 2MASS catalog; these are potential candidate SiO maser sources, and 
 from past experience, half of these objects exhibit SiO masers. 
The results are shown in the last two columns of table \ref{tab:2},  listing
the number of NIR counterparts satisfying the above conditions within 34$'$ (1 square degree) of the GC 
(between parentheses, the number of MSX sources above 1 Jy in the same area),
and the probability of contamination,
finding more than one such object within $r_{\rm SiO}$
toward the globular cluster with a SiO detection rate of 50\%.  These numbers are quite consistent with
the value obtained from the above estimate based on the known SiO maser sources in the bulge.

In addition, the radial velocities of the bulge SiO maser sources
spread over approximately $\pm 200$
km s$^{-1}$. The chance of an accidental coincidence
of a bulge SiO source both in position and velocity is thus
considered to be quite small
unless the motions of the bulge GCs are dominated by
bulge-bar star kinematics, though this is somewhat dubious;
e.g., \citet{Cote-1999}.
While that in Liller~1 is probably in fact an
accidental coincidence, it is highly unlikely that all of our
detections were by chance.

It is also interesting to note that the present detections of SiO masers are
limited to the bulge/disk GCs. As mentioned in section \ref{sec:Intro},
a previous massive SiO maser search with the VLA
failed to find SiO maser emission in any halo GCs.
Furthermore, mass-losing AGB candidates, i.e.,
the very red stars found in our NIR observations and/or those identified
in the MSX/IRAS surveys, are only found in bulge clusters.
These facts indicate that the bulge GCs
have AGB stellar populations that differ in their characteristics
from those in the halo.
A possible 
explanation is that the bulge GCs are relatively
metal-rich ($\sim 1/10$ to 1 solar abundance; see table \ref{tab:2}).
The molecular abundances, especially for SiO, are enhanced
in a metal-rich environment.

\subsection{Bolometric Magnitude\label{sec:Mbol}}

Because
the peaks of the spectral energy distribution of
mass-losing AGB stars, such as those discussed here, lie
in the region of the NIR and MIR,
their bolometric magnitudes can be estimated by combining our NIR photometric
data with MIR catalogs, such as IRAS and MSX \citep{Egan-2003}. 
For the newly discovered variables,
we adopted mean magnitudes from our NIR light curves.
They all show large amplitudes that are typical for mass-losing stars.
For the other two variables that are too bright for us to monitor,
we adopted their magnitudes from the 2MASS All
Sky Data Release \citep{Cutri-2003}.
Comparing the magnitude of NGC~6171 V1 with that 
reported by \citet{Feast-1994}, it is probably not far from its mean magnitude.
The adopted magnitudes are listed in table \ref{tab:3},
which also indicates the corresponding MSX source and/or
IRAS point source, if any.
The bolometric magnitudes, $m_{\rm bol}$ were calculated by integrating
under trapezoidal lines drawn in $\log \nu$--$\nu F_\nu$ diagrams.
While we ignored the contributions from shorter wavelengths,
we applied Rayleigh--Jeans curves for the longer ones.
The uncertainties are expected to be
about 0.2 mag, mainly arising from the uncertainties of the mean-magnitude
determinations in the NIR and the flux density errors in the MIR.
In table \ref{tab:3} the results are also listed  along
with the derived absolute bolometric magnitudes obtained by correcting
for the assumed cluster distance moduli and reddening values.


\begin{table*}
  \caption{Photometric properties of SiO maser sources.}\label{tab:3}
 \begin{center}
  \begin{tabular}{lrrrcccc}
  \hline\hline
Object& $J$& $H$&
$K_{\rm s}$& MSX6C&  IRAS PSC& $m_{\rm bol}$& derived $M_{\rm bol}$ 
\\
\hline
  Liller~1 V1 & 14.70 & 11.80 & 9.00 & G354.8792$-$00.1798 & --- & 11.0 & $-$3.8 \\
  NGC~6171 V1 & 6.02 & 5.16 & 4.54 & --- & 16296$-$1305 & 7.4 & $-$6.6 \\
  Pal~6 V1    &  ---  & 12.00 & 9.00 & G002.0765+01.7380 & 17406$-$2614 & 10.4 & $-$3.5  \\
  Pal~10 V2   & 6.79 & 5.58 & 4.87 & G052.4153+02.7613 & 19156+1828 & 7.4 & $-$6.4  \\
  Terzan~5 V5 & 9.90  & 8.20  & 6.90  & G003.8381+01.6895 & --- & 9.3 & $-$5.6 \\
  Terzan~12 V1& 9.30  & 7.70  & 6.50  & G008.3630$-$02.0909 & 18092$-$2244 & 9.0 & $-$4.3 \\
  \hline
 \end{tabular}

  \end{center}
\end{table*} 

The derived $M_{\rm bol}$ values should be lower than those of the AGB tips,
which depend on the age and chemical composition. \citet{Aaronson-1985} showed
that the younger AGB stars evolve upwards in luminosity.
They consider that the tip luminosity for metal-rich GCs is
about $-4.5$ mag at an age of about 10 Gyr. Only Pal~6 V1 and Terzan~12
have a $M_{\rm bol}$ value consistent with this limit,
i.e.,  lower than $-4.5$.
On the other hand, the merger of binary systems or direct collision
can produce massive objects that would reach higher luminosity
\citep{Bailyn-1995}.
This possibility is also mentioned by \citet{Feast-1994}.
This is related to a problem concerning planetary nebulae (PNe) in GCs.
So far, four PNe are known to exist in GCs
(Ps-1 in M~15, \cite{Pease-1928}; GJJC-1 in M~22, \cite{Gillett-1989};
JaFu-1 in Pal~6 and JaFu-2 in NGC~6441, \cite{Jacoby-1997}).
While this number is less than expected based on the assumption
that every dying star produces a PN,
current models predict that the masses of white dwarfs in GCs 
are too small to produce PNe (see \cite{Jacoby-1997}; \cite{Alves-2000},
and references therein). These authors suggested that
only special objects, whose masses are augmented in close-binary interaction,
can become PNe. Because the variable stars with heavy mass-loss 
have a lifetime roughly equal to that of planetary nebulae,
it seems natural to expect the same number of them,
both of which are considered to be products of merged binaries.
Since blue stragglers are common in GCs \citep{Piotto-2004}
and also with the recent discoveries that they contain
numerous low-mass X-ray binaries and 
millisecond pulsars \citep{Lyne-2000},
it is strongly suggested that close-binary formations and 
merger events occur relatively often in them \citep{Pooley-2003}.
With these indications and our new detection of masers with high luminosities
(NGC~6171 V1, Pal~10 V2, and Terzan~5 V5),
we support the idea of a merged binary by \citet{Feast-1994}.
It may be slightly premature, however, at this stage to discuss 
their statistics and 
formation rates before further complete surveys are made.

\section{Conclusion}
Among the six SiO maser sources that we detected in the present survey,
five, with the exception of Liller~1 V1, stand as  potential candidates for GC members.
Among them, the variables toward NGC 6171 and Terzan 5 are highly likely GC members,
though those toward Pal 6, Pal 10,  and Terzan 12, are somewhat debatable.
The membership must be approved (or disproved) much more firmly
in future observations, because
some uncertainty  remains as to the coincidence
of the maser velocities with those of the clusters
due to a lack of precision in the optical velocity measurements of GCs.
These maser sources are dust-enshrouded AGB stars that were not
expected in GCs in the past. Some of them have luminosities that are too high
to fit the standard evolutionary tracks of low-mass stars, and may have evolved
from binary mergers. On the other hand, SiO sources toward Pal~6 and Terzan~12 have
luminosities that are consistent with the normal stellar evolution of old stars, 
though they might also be special objects at a certain stage of close-binary
interaction.

\

The authors are grateful to the staff members of Nobeyama Radio Observatory 
for supporting the 45-m telescope operations.
They also  thank Dr. I.~S. Glass for his thorough reading of the manuscript,
Dr. M.~W. Feast for useful comments, and 
the IRSF/SIRIUS team for near-infrared observations.
Some of the near-infrared data were obtained by
Dr. M.~W. Feast, Dr. J.~W. Menzies, S. Nishiyama, and D. Baba.
The IRSF/SIRIUS project was initiated and supported by Nagoya University,
National Astronomical Observatory of Japan, and University of Tokyo
in collaboration with South African Astronomical Observatory
under a financial support of Grant-in-Aid for Scientific Research
on Priority Area (A) No. 10147207 and 10147214 of the Ministry of Education,
Culture, Sports, Science and Technology.

\end{document}